\documentclass[12pt]{iopjournal}

\usepackage{amsmath,amssymb}
\usepackage{bm}
\usepackage{braket}
\usepackage{chemformula}
\usepackage{placeins}
\usepackage[numbers,sort&compress]{natbib}
\usepackage{ragged2e}

\newcommand{\figsubref}[2]{\hyperref[#1]{\ref*{#1}(#2)}}
\usepackage[no-math]{fontspec}
\newfontfamily\RoPPTitleFace{TeXGyreTermesX-Bold.otf}
\newcommand{\RoPPTitleFont}{%
  \RoPPTitleFace\fontsize{16}{16}\selectfont
}

\begin{document}

\title{{\RoPPTitleFont
Bulk $g$-wave altermagnetism induced mode selectivity in skyrmion tubes}}
\author{Yuchi Yang and Yusheng Hou$^{*}$}\par
\affil{Guangdong Provincial Key Laboratory of Magnetoelectric Physics and Devices, Institute of Neutron Science and Technology, School of Physics, 
Sun Yat-Sen University, Guangzhou 510275, China}\par

\email{houysh@mail.sysu.edu.cn}\par

\begin{abstract}
    
\justifying
\setlength{\emergencystretch}{2em}

Altermagnetic skyrmions have attracted interest in spintronics for their symmetry-dependent shapes and motion. Although skyrmions in $d$-wave altermagnets have been studied, the formation of skyrmions in bulk $g$-wave altermagnets remains unclear. Using a frustrated atomistic spin model, we find skyrmion tubes in bulk $g$-wave altermagnets are distinct from their $d$-wave counterparts. The $g$-wave symmetry has no effect on their static configurations and energies, which are consistent with those of the corresponding antiferromagnetic model without altermagnetism. Its influence instead emerges in the energy curvature. Indeed, the altermagnetism selectively modifies the curvature, with the selectivity embodied in both the tube axis and in-plane angular structure. For the former, the $g$-wave contribution to the Hessian vanishes at the axial nodes, $q_zc=0,\pi$, and allows modifications only away from these nodes, revealing an out-of-plane signature of altermagnetism in the skyrmion modes. For the latter, the $g$-wave symmetry acting on a mode component of angular order $m$ selectively produces components of orders $m\pm3$, while the breathing-like mode offers an obvious manifestation in its threefold deformation. These features of selectivity also extend to dynamic eigenmodes. Our results provide a comprehensive framework for understanding bulk $g$-wave altermagnetic skyrmion tubes, offering new insights into the $g$-wave altermagnetic symmetry.

\end{abstract}

\justifying
\setlength{\emergencystretch}{2em}

\section{Introduction}

Magnetic skyrmions are topologically nontrivial spin textures characterized by a nonzero topological charge. Since their first experimental observation in chiral magnets, magnetic skyrmions have attracted considerable attention~\cite{2009_Skyrmion_Lattice_in_a_Chiral_Magnet_for_skyrmions_observation,2010_Real-space_observation_for_skyrmion_observation,2012_Observation_of_Skyrmions_for_skyrmion_observation}. Their nanoscale size, long lifetime, and current-driven motion at low current densities~\cite{2013_Topological_properties_and_dynamics_of_magnetic_skyrmions_for_skyrmions_review,2017_Magnetic_skyrmions_advances_for_skyrmion_review,2021_Magnetic_Skyrmion_Materials_for_skyrmion_review} make them promising candidates for advanced spintronic applications, such as racetrack memory~\cite{2013_Skyrmions_on_the_track,2014_A_strategy_for_skyrmion_racetrack_memories} and neuromorphic computing~\cite{2020_Reservoir_Computing_with_Random_Skyrmion_Textures,2018_Magnetic_Skyrmion_as_a_Nonlinear_Resistive_Element,2020_Skyrmion-based_artificial_synapses_for_neuromorphic_computing}. Although commonly described as two-dimensional (2D) objects, skyrmions can extend along the $z$ axis to form tubes in three-dimensional (3D) magnets, which in the simplest case consist of the uniform stacking of a 2D skyrmion spin texture~\cite{2013_Skyrmion_lattice_phase_for_simulations_of_science,2022_Direct_visualization_of_the_three-dimensional_shape_for_3D,2022_Unveiling_the_three-dimensional_magnetic_texture_for_3D,2020_Real-space_imaging_of_confined_magnetic_skyrmion_tubes_for_3D}. Indeed, the skyrmion lattice originally observed in bulk \ch{MnSi} consisted of skyrmion tubes extending along the applied magnetic field~\cite{2009_Skyrmion_Lattice_in_a_Chiral_Magnet_for_skyrmions_observation,2013_Skyrmion_lattice_phase_for_simulations_of_science}. The 3D structure of these tubes also plays an important role in skyrmion creation, annihilation, and motion under external stimuli~\cite{doi:10.1126/sciadv.aat1115,2020_Real-space_imaging_of_confined_magnetic_skyrmion_tubes_for_3D,2023_Three-dimensional_structure_of_hybrid_magnetic_for_3D,Kagawa2017}.

Skyrmions have been extensively studied in ferromagnets (FM)~\cite{2013_Topological_properties_and_dynamics_of_magnetic_skyrmions_for_skyrmions_review,2017_Magnetic_skyrmions_advances_for_skyrmion_review}, while their counterparts in intrinsic and synthetic antiferromagnets (AFM) have also attracted interest~\cite{2020_Room-temperature_stabilization_of_antiferromagnetic_skyrmions_for_AFM,2022_Emergence_of_zero-field_non-synthetic_single_for_AFM}. The appeal of AFM skyrmions is partly due to their current-driven motion without the skyrmion Hall effect (SkHE)~\cite{2016_Magnetic_bilayer-skyrmions_without_skyrmion_Hall_effect_for_AFM,2016_Static_and_Dynamical_Properties_of_Antiferromagnetic_Skyrmions_for_AFM,2024_Fast_current-induced_skyrmion}. More recently, altermagnetism (AM) has been identified as a distinct category of collinear magnetic order beyond FM and AFM~\cite{beyond_conventional,Emerging_Research_Landscape_of_Altermagnetism,Altermagnetism---A_NewPunch——Line}. Together with momentum-dependent spin splitting in reciprocal space, AMs are characterized by compensated antiparallel magnetic order in real space. Their opposite sublattices are related by crystal rotations rather than translation or inversion as in conventional AFMs. By combining the advantages of both FMs and AFMs, AMs offer promising opportunities for spintronics~\cite{2025_Altermagnets_as_a_new_class_of_functional_materials,2026_Altermagnetic_spintronics}. In a $d$-wave AM model, skyrmions can form elliptical pairs with a $90^\circ$ interlayer orientation owing to the anisotropic exchange interactions~\cite{2024_Skyrmion_Hall_Effect_in_Altermagnets_for_skyrmion_hall,dwave_skymions_with_splitting}. Meanwhile, monolayer $\mathrm{V}_2\mathrm{SeTeO}$ has been proposed as a candidate system hosting altermagnetic skyrmions~\cite{2026_Altermagnetic_skyrmions_in_two-dimensional_lattices_for_AM}. Beyond these structural features, theoretical studies have predicted the presence of an anisotropic SkHE~\cite{2024_Skyrmion_Hall_Effect_in_Altermagnets_for_skyrmion_hall,2025_Spin-Transfer_Torque_in_Altermagnets_for_AM,2026_Altermagnetic_skyrmions_in_two-dimensional_lattices_for_AM}. These results illustrate the reflection of AM symmetry in both the structure and motion of skyrmions.

However, studies of AM skyrmions have mainly focused on $d$-wave AMs, leaving their counterparts in bulk $g$-wave AMs unexplored despite their considerable experimental attention. Among these bulk $g$-wave AMs, metallic \ch{CrSb} provides a well-characterized example, combining a high N\'eel temperature with large spin splitting near the Fermi level~\cite{Reimers2024,2024_Large_Band_Splitting_in_g-Wave_Altermagnet_CrSb_for_g_spin_splitting,2024_Observation_of_Spin_Splitting_for_CrSb}. Angle-resolved photoemission spectroscopy and quantum oscillation measurements have established the bulk $g$-wave symmetry of the splitting~\cite{2025_Three-dimensional_mapping_of_CrSb,2026_3D_bulk-resolved_g-wave_altermagnetic_order_parameter_in_CrSb_for_g_spin_splitting}. Semiconducting \ch{MnTe} is another $g$-wave AM with similar symmetry~\cite{2023_Altermagnetism_in_MnTe,2024_Broken_Kramers_Degeneracy_in_Altermagnetic_MnTe,2024_Observation_of_splitting_in_MnTe}. For these hexagonal AMs, the spin splitting takes the form of $k_zk_y(3k_x^2-k_y^2)$~\cite{2024_Minimal_models_for_altermagnetism}, including an out-of-plane $k_z$ factor. Previous work on the $d$-wave case has suggested the relevance of spin splitting symmetry to skyrmions, whose shapes provide a macroscopic realization of the AM symmetry~\cite{dwave_skymions_with_splitting}. The additional $k_z$ dependence in bulk $g$-wave AMs such as \ch{CrSb} and \ch{MnTe} opens a new avenue for exploring the out-of-plane signatures of AM in skyrmion textures.

In this work, we investigate the influence of bulk $g$-wave altermagnetism on skyrmion tubes using an atomistic spin model that captures the AM symmetry of \ch{CrSb} and \ch{MnTe}. In this model, N\'eel skyrmion, Bloch skyrmion, and antiskyrmion tubes share the same energy. Moreover, in contrast to the elliptical $d$-wave skyrmion shapes~\cite{2024_Skyrmion_Hall_Effect_in_Altermagnets_for_skyrmion_hall,dwave_skymions_with_splitting}, the relaxed textures exhibit no in-plane anisotropy induced by the $g$-wave AM symmetry. This symmetry contribution cancels in the static configurations, leaving both the textures and their energies unchanged. This static cancellation motivates a further analysis of the energy curvature through the Hessian matrix, which is closely related to localized magnon modes~\cite{2018_Localized_spin_waves,2023_Eigenmodes_of_magnetic_skyrmion_lattices_for_dynamics_modes,2020_Spin_waves_in_skyrmionic} and the lifetime~\cite{2012_HTST_for_lifetime,2018_Energy_surface_and_lifetime,2018_Thermal_stability_of_metastable_magnetic_skyrmions_for_lifetime,2019_Skyrmion_lifetime_in_ultrathin_films} of skyrmions. We find that the bulk $g$-wave symmetry selectively modifies the Hessian eigenmodes away from the axial nodes $q_zc=0,\pi$, while its contribution vanishes at these axial nodes. Actually, this axial selectivity arises from the $\sin(q_zc)$ dependence of the $g$-wave contribution to the Hessian and therefore depends on the axial wave vector. Furthermore, the in-plane angular spectra reveal a selective shift of $\Delta m=\pm3$ between the modes themselves and their $g$-wave AM symmetry action. This is particularly evident in the threefold deformation of the breathing-like mode. Finally, we find that the axial and in-plane angular selectivity is inherited by the dynamic eigenmodes, providing a potential experimental signature of bulk $g$-wave AM in skyrmion tubes.

\section{Model and methods}

\begin{figure}[htbp]
    \centering
    \includegraphics[width=0.95\linewidth]{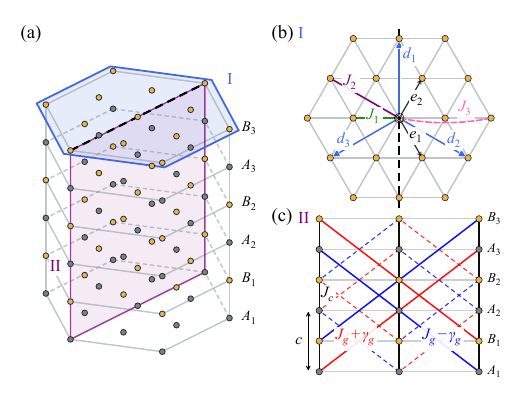}
    \caption{Schematic of the bulk $g$-wave structure. (a) 3D view of the structure. Note that the axial distance is slightly enlarged for clarity. Gray and yellow sites denote sublattices A and B. (b) Triangular lattice in plane {\uppercase\expandafter{\romannumeral1}}, with primitive vectors $e_1$ and $e_2$. (c) Vertical section through plane {\uppercase\expandafter{\romannumeral2}}.}
    \label{fig:structure}
\end{figure}

We consider a classical spin model in Fig.~\figsubref{fig:structure}{a} with competing intralayer exchange interactions. As illustrated in Fig.~\figsubref{fig:structure}{b}, $J_1$, $J_2$, $J_3$ are the intralayer nearest-neighbor, second-nearest-neighbor and third-nearest-neighbor exchange strengths, respectively. In Fig.~\figsubref{fig:structure}{c}, the magnetic lattice consists of alternating triangular layers belonging to sublattices $A$ and $B$. The repeat distance between equivalent layers along the $z$ axis is $c$, with adjacent $A$ and $B$ layers separated by $c/2$. $J_c$ is the interlayer antiferromagnetic exchange strength, connecting sublattices $A$ and $B$. The inequivalent diagonal bonds $J_g\pm\gamma_g$ denote anisotropic interlayer exchange, which connect sites on the same sublattice in layers separated by $c$. The sign in $J_g\pm\gamma_g$ depends on both the sublattice and the bond orientation as shown in Fig.~\figsubref{fig:structure}{c}, encoding the bulk $g$-wave AM symmetry~\cite{g-wave_model1,g-wave_model2}.

The atomistic model Hamiltonian is in the following form: 
\begin{equation}
\begin{aligned}
\mathcal{H}=&-J_1
\sum_{\langle i,j\rangle}\mathbf{S}_i\cdot\mathbf{S}_j
-J_2\sum_{\langle\langle i,j\rangle\rangle}\mathbf{S}_i\cdot\mathbf{S}_j
-J_3\sum_{\langle\langle\langle i,j\rangle\rangle\rangle}\mathbf{S}_i\cdot\mathbf{S}_j-J_c\sum_{\langle i,j\rangle_\mathrm{AB}}\mathbf{S}_i\cdot\mathbf{S}_j
\\ &
-J_g\left(\sum_{\langle i,j\rangle_\mathrm{up}}+\sum_{\langle i,j\rangle_\mathrm{dw}}\right)
\mathbf{S}_i\cdot\mathbf{S}_j+\gamma_g\left(\sum_{\langle i,j\rangle_\mathrm{up}}-\sum_{\langle i,j\rangle_\mathrm{dw}}\right)\xi_i\mathbf{S}_i\cdot\mathbf{S}_j-K\sum_i\left(S_i^z\right)^2,
\end{aligned}
\label{eq:spin Hamiltonian}
\end{equation}
where $K$ is the magnetic anisotropy coefficient. The first three exchange sums run over the intralayer neighbor shells defined above, and $\langle i,j\rangle_\mathrm{AB}$ denotes the nearest-neighbor bonds between $A$ and $B$ sublattices. $\langle i,j\rangle_\mathrm{up}$ and $\langle i,j\rangle_\mathrm{dw}$ include diagonal bonds on both sublattices with displacement vectors $d_p+ce_z$ and $d_p-ce_z$, respectively, where $p=1,2,3$. The in-plane displacement vectors are $d_1=-e_1+e_2$, $d_2=2e_1+e_2$, and $d_3=-e_1-2e_2$, as indicated by blue arrows in Fig.~\figsubref{fig:structure}{b}. The sublattice factor is $\xi_i=-1$ for $i\in A$ and $\xi_i=+1$ for $i\in B$. The Dzyaloshinskii--Moriya interaction (DMI) is neglected here since its influence on the AM splitting is qualitatively small~\cite{g-wave_model1}. To investigate the role of the $g$-wave symmetry alone, we separate the Hamiltonian in Eq.~\ref{eq:spin Hamiltonian} into
\begin{equation}
    \mathcal{H}=\mathcal{H}_a+\gamma_g\mathcal{H}_g,
    \label{eq:hamiltonian_carrier_split}
\end{equation}
where
\begin{equation}
    \gamma_g\mathcal{H}_g=\gamma_g\left(\sum_{\langle i,j\rangle_\mathrm{up}}-\sum_{\langle i,j\rangle_\mathrm{dw}}\right)\xi_i\mathbf{S}_i\cdot\mathbf{S}_j.
    \label{eq:g-wave carrier}
\end{equation}
Here, $\gamma_g\mathcal{H}_g$ is the carrier of the $g$-wave AM symmetry~\cite{g-wave_model1,g-wave_model2}, and $\mathcal{H}_a$ denotes the remaining terms.

The spin configurations are obtained using the atomistic simulation framework Spirit~\cite{spirit}. The main calculations are performed on a $32\times32\times4$ supercell with periodic boundary conditions along all three directions. The parameters are set to $J_1=1~\mathrm{meV}$, $J_2/J_1=-0.28$, $J_3/J_1=-0.06$, $J_c/J_1=-0.74$, $J_g=0$, $\gamma_g/J_1=-0.10$ and $K/J_1=0.09$ in the absence of an external magnetic field. Initial states are constructed by stacking a standard 2D N\'eel, Bloch, or antiskyrmion texture along $z$, with opposite spin directions on adjacent $A$ and $B$ layers. This construction yields straight skyrmion tubes extending along the $z$ direction. All spins are then relaxed using the velocity projection method with a spin-force convergence threshold of $10^{-8}$ in Spirit~\cite{spirit}. Additional relaxation tests using $64\times64\times4$ supercells, weakly perturbed initial states, and another equation solver also yield identical relaxed textures. The relaxed antiskyrmion configuration obtained from the unperturbed initial state in the $32\times32\times4$ supercell serves as the reference state for further analysis. Figure~\figsubref{fig:three texture and corresponding}{c} shows one layer of this tube.

Within the harmonic approximation, we expand the Hamiltonian around the relaxed spin configuration $S^{(0)}$ as~\cite{2018_Localized_spin_waves}
\begin{equation}
    \mathcal{H}\simeq\mathcal{H}_0+\frac{1}{2}(\delta S)^TH\delta S,
    \label{eq:hessian expansion}
\end{equation}
where $\mathcal{H}_0$ is the energy of the relaxed configuration, and $\delta S$ collects the two local transverse spin deviation components of each site. The Hessian matrix $H$, obtained from Spirit~\cite{spirit}, describes the curvature of the potential energy surface with respect to the transverse spin deviations. Diagonalizing $H$ yields the eigenvalues $\lambda_n$ and eigenvectors $u_n$.

To evaluate the similarity of Hessian eigenmodes obtained from different calculations, we calculate their overlap as~\cite{2026_eigenmode_following_method_for_mode_corresponding}
\begin{equation}
P_s=\left|v^{\dagger}u\right|,
\label{eq:single_mode_overlap}
\end{equation}
where $u$ and $v$ are normalized eigenvectors from different calculations. A value of $P_s$ close to unity indicates agreement between two eigenmodes. For nearly degenerate modes, comparing individual eigenvectors alone may not reliably establish mode similarity. Therefore, we compare the whole subspaces spanned by the group of modes. For two groups of $r$ modes, represented by orthonormal eigenvectors $\{u_n\}$ and $\{v_a\}$, we calculate their subspace overlap,
\begin{equation}
P_g=\sqrt{\frac{1}{r}\sum_{a=1}^{r}\sum_{n=1}^{r}\left|v_a^{\dagger}u_n\right|^2},
\label{eq:mode_subspace_overlap}
\end{equation}
where $\sum_{n=1}^{r}\left|v_a^{\dagger}u_n\right|^2$ measures how fully $v_a$ lies within the subspace spanned by $\{u_n\}$. Therefore, a value of $P_g$ close to unity indicates the agreement between the two subspaces, regardless of the choice of orthonormal basis. For convenience, both $P_s$ for isolated modes and $P_g$ for nearly degenerate modes are denoted by $P$ in the following discussion.

\section{Results}

\subsection{Static Cancellation and Energy Curvature}

\begin{figure}[htbp]
    \centering
    \includegraphics[width=1.0\linewidth]{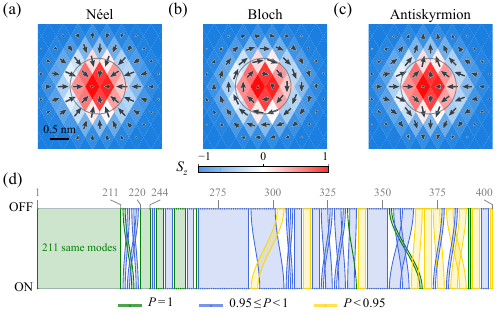}
    \caption{Relaxed spin textures and correspondence between Hessian eigenmodes with and without the $g$-wave carrier. (a)--(c) Spin textures in a layer of sublattice $A$ for the relaxed N\'eel skyrmion, Bloch skyrmion, and antiskyrmion tubes, respectively. Arrows show the in-plane spin components. Gray lines mark the $S_z=0$ contours, extracted using piecewise cubic interpolation. (d) Correspondence between the lowest 400 Hessian modes calculated with the $g$-wave carrier switched off and on. Green indicates agreement of both eigenvalues and eigenvectors or eigensubspaces, with $P=1$. The remaining correspondences are determined by maximizing the overlap. Blue and yellow indicate $0.95\leq P<1$ and $P<0.95$, respectively. The scale is compressed in the two long green intervals.}
    \label{fig:three texture and corresponding}
\end{figure}

We first examine the localized spin textures relaxed from three initial configurations: a N\'eel skyrmion, a Bloch skyrmion, and an antiskyrmion tube. As shown in Figs.~\figsubref{fig:three texture and corresponding}{a}--\figsubref{fig:three texture and corresponding}{c}, the relaxed textures retain their respective N\'eel, Bloch, and antiskyrmion character with the same size. Moreover, the three relaxed textures have the same total energy. This degeneracy can be attributed to the symmetry of the inversion-symmetric model without DMI and dipole-dipole interaction. In this model, a global spin rotation around the $z$ axis connects the N\'eel and Bloch textures, and the transformation $S_y\rightarrow -S_y$ reverses the in-plane winding and connects a skyrmion to an antiskyrmion without changing the energy~\cite{2016_Ginzburg-Landau_theory_for_skyrmions_for_three_energy_degenerate,2017_Skyrmion_dynamics_in_a_frustrated_ferromagnetic_for_three_energy_degenerate}. Thus, the three textures differ in their in-plane spin arrangements while sharing the same $S_z$ profile and energy.

Beyond the energy degeneracy, the relaxed textures in Figs.~\figsubref{fig:three texture and corresponding}{a}--\figsubref{fig:three texture and corresponding}{c} show no in-plane anisotropy associated with the $g$-wave exchange pattern. Meanwhile, they coincide with those obtained from the AFM model without $g$-wave altermagnetism. This contrasts with the results in the $d$-wave AM model, where anisotropic intralayer exchange produces elliptical skyrmions elongated and compressed along mutually perpendicular directions on the two sublattices~\cite{dwave_skymions_with_splitting}. To further determine the effect of the carrier, we repeat the relaxation calculations for $\gamma_g/J_1$ ranging from $-0.20$ to $0.20$ in steps of $0.025$. The result with $\gamma_g/J_1=0$ corresponds to the antiferromagnetic case without AM symmetry. However, both the total energy and spin profile remain the same across the tested $\gamma_g$, indicating a negligible effect of $\gamma_g$ on the static configuration and energy. Indeed, this invariance arises from an axial cancellation of the carrier contribution in Eq.~\ref{eq:spin Hamiltonian}. For a given in-plane displacement $d_p$, the two diagonal bonds with displacements $d_p\pm ce_z$ connect a given spin to two identically oriented spins along the tube axis. Since the paired diagonal bonds in $\gamma_g\mathcal{H}_g$ have opposite signs, their contributions cancel, leaving the static configurations and energies unchanged with different $\gamma_g$. 

However, it is still unresolved how the $g$-wave carrier $\gamma_g\mathcal{H}_g$ affects the skyrmion tubes. Therefore, beyond the static energy, we turn to the Hessian matrix $H$, which governs the quadratic energy term in Eq.~\ref{eq:hessian expansion}. We compare its eigenmodes with the carrier switched off, $\gamma_g/J_1=0$, and on, $\gamma_g/J_1=-0.10$. As shown in Fig.~\figsubref{fig:three texture and corresponding}{d}, even the lowest 211 Hessian modes remain unchanged, with the same eigenvalues and eigenvectors. It is not until modes 212 and 213 in the carrier-on spectrum, corresponding to modes 214 and 215 in the carrier-off spectrum, that a difference emerges. The eigenvalues of this pair shift, while their overlap remains close to unity ($P=0.998$). Among the lowest 400 modes in Fig.~\figsubref{fig:three texture and corresponding}{d}, 254 remain unchanged, and the eigenvalues of the remaining 146 change. However, most of the affected modes retain high overlaps, $P\geq0.95$, indicating that the corresponding mode subspaces remain closely aligned. These results show that the $g$-wave carrier $\gamma_g\mathcal{H}_g$ leaves the static textures and energies unchanged, yet modifies the energy curvature along some modes.

\subsection{Axial Selectivity of the Carrier}

\begin{figure}[htbp]
    \centering
    \includegraphics[width=1\linewidth]{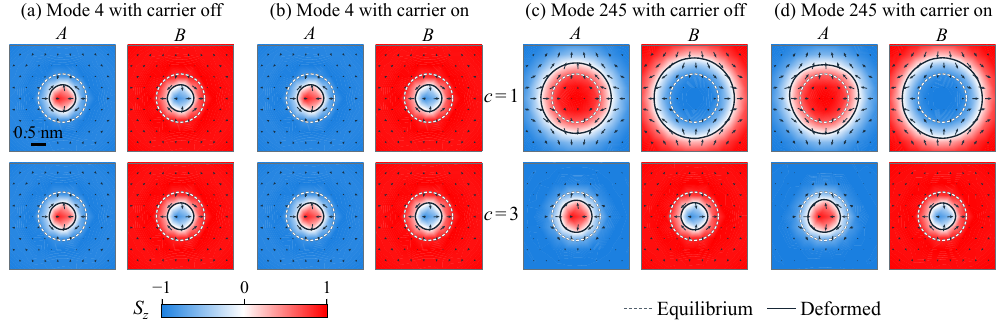}
    \caption{Spin texture deformations associated with Hessian modes 4 and 245. (a), (b) Deformations of mode 4 with the carrier off and on, respectively. (c), (d) Corresponding results for mode 245. Within each panel, the left and right columns show sublattices $A$ and $B$, respectively. The top and bottom rows correspond to the first and third unit cells along $z$, respectively. The dotted and solid lines denote the $S_z=0$ contours of the equilibrium and deformed configurations, respectively.}
    \label{fig:two modes with on/off}
\end{figure}

The preceding comparison shows that the $g$-wave carrier affects some Hessian modes while raising another question of what distinguishes the changed modes from those remaining unchanged. As representative examples, we compare mode 4, an unchanged breathing mode, with mode 245, a breathing-like mode affected by the carrier. As shown in Fig.~\ref{fig:two modes with on/off}, the two modes exhibit different deformation patterns along the tube axis, both with and without the carrier. In Figs.~\figsubref{fig:two modes with on/off}{a} and \figsubref{fig:two modes with on/off}{b}, mode 4 produces the same radial deformation pattern in the first and third unit cells along $z$. In contrast, mode 245 exhibits expansion in the first unit cell and shrink in the third, as shown in Figs.~\figsubref{fig:two modes with on/off}{c} and \figsubref{fig:two modes with on/off}{d}. Moreover, as shown in Fig.~S1 of the Supplementary Material (SM)~\cite{SupplementaryMaterial}, the deformation profiles across all four unit cells further support the axial difference between modes 4 and 245.

To quantify this difference, we introduce a normalized interlayer correlation
\begin{equation}
    C_{ab}=\frac{\sum_i t_{a,i}\cdot t_{b,i}}{\sqrt{\sum_i|t_{a,i}|^2\cdot\sum_i|t_{b,i}|^2}}.
    \label{eq:interlayer correlation}
\end{equation}
Here, $t_{a,i}$ is the spin perturbation at site $i$ in unit cell $a$, with $i$ running over all sites of both sublattices $A$ and $B$. Therefore, the interlayer correlation $C_{ab}$ measures the similarity between the spin perturbation patterns in unit cells $a$ and $b$. For mode 4, $C_{12}=C_{13}=C_{14}=1$, suggesting a common perturbation pattern across all four unit cells. In contrast, mode 245 satisfies $C_{13}=C_{24}=-1$ and $C_{12}=-C_{14}$. This suggests that the perturbation pattern reverses between unit cells separated by $2c$. Remarkably, this distinction is not limited to these two representative modes, but extends to all 400 modes in Fig.~\figsubref{fig:three texture and corresponding}{d}, without exception. The spin perturbation patterns of all 254 unchanged modes repeat every $c$ like mode 4. The remaining 146 changed modes follow the axial behavior of mode 245, reversing sign after a translation of $2c$ and recovering their original form after $4c$. Therefore, the modification arising from the carrier is possibly related to axial dependence.

As discussed above, these two groups of modes exhibit distinct periodicities along the $z$ axis. Since the relaxed background repeats with period $c$, we can describe the axial variation of the spin perturbations using the wave vector $q_z$~\cite{2019_Kelvin_modes_of_a_skyrmion_line_for_qz}. Therefore, for a mode with a wave vector $q_z$, the eigenvector can be expressed in Bloch form as
\begin{equation}
    u_n=
    \begin{pmatrix}
        u_{n,1} \\
        u_{n,2} \\
        u_{n,3} \\
        u_{n,4} \\
    \end{pmatrix}=\frac{1}{2}
    \begin{pmatrix}
        e^{iq_zc}u^q_{n} \\
        e^{2iq_zc}u^q_{n} \\
        e^{3iq_zc}u^q_{n} \\
        e^{4iq_zc}u^q_{n} \\
    \end{pmatrix},
    \label{eq:reconstruction of eigenvectors}
\end{equation}
where $u_n$ is the eigenvector of four full cells, $u_{n,a}$ is its component in unit cell $a$, and $u^q_{n}$ is the single-cell eigenvector at $q_z$. The factor $1/2$ ensures the normalization of both single-cell $u^q_{n}$ and four-cell $u_n$. For our four-cell system with periodic boundary conditions, the allowed wave vectors satisfy $q_zc=0,\pm\pi/2,\pi$. Accordingly, the common perturbation pattern of the unchanged modes may correspond to the $q_zc=0$ sector, with $u_{n,1}=u_{n,2}=u_{n,3}=u_{n,4}$. The sign reversal after two unit cells in the changed modes may instead correspond to the $q_zc=\pm\pi/2$ sectors, with $e^{2iq_zc}=-1$ and $e^{4iq_zc}=1$.

Moreover, the Hessian matrix can be reconstructed at different $q_z$ within the Bloch representation. Considering that the Hamiltonian in Eq.~\ref{eq:spin Hamiltonian} contains only intracell terms and couplings between adjacent unit cells along $z$, the reconstructed matrix is in the following form
\begin{equation}
    H^q=B_0+e^{iq_zc}B_{+1}+e^{-iq_zc}B_{-1},
    \label{eq:reconstructed matrix}
\end{equation}
with the detailed derivation provided in Sec.~S2 of the SM~\cite{SupplementaryMaterial}. Here, $B_0$ contains the Hessian contributions within one unit cell, while $B_{+1}$ and $B_{-1}$ couple it to the neighboring cells at $+c$ and $-c$, respectively. These three matrices can be extracted from the original Hessian matrix $H$. The Bloch phase factors generally make $H^q$ complex, and diagonalizing the complex Hessian matrix $H^q$ therefore yields complex single-cell eigenvectors $u_n^q$. Nevertheless, its eigenvalues $\lambda_n^q$ are real because the reconstructed matrix $H^q$ is Hermitian. This Hermiticity stems from $B_{+1}=B_{-1}^{T}$ given that the original Hessian $H$ is real and symmetric. Meanwhile, the matrices $B_{0}$ and $B_{\pm1}$ are also real, giving $H^{-q}=(H^q)^*$. Taken together, a mode at $q_z$ thus has a complex-conjugate counterpart at $-q_z$ with an identical eigenvalue. This reconstruction describes the corresponding modes through a single-cell Hessian $H^q$, and connects the observed axial repetition patterns in Fig.~\ref{fig:two modes with on/off} to different $q_z$ sectors.

To examine the reliability of our reconstruction, we compare the original eigenmodes directly from the four-cell Hessian $H$ using Spirit~\cite{spirit}, with those reconstructed from the single-cell Hessian $H^q$ in Eq.~\ref{eq:reconstructed matrix}. For single-cell eigenmodes $u_n^q$ at wave vector $q_z$, we use Eq.~\ref{eq:reconstruction of eigenvectors} to extend each $u_n^q$ to a full four-cell eigenvector. Considering that the modes at opposite wave vectors form a conjugate pair for $q_zc=\pm\pi/2$, the real and imaginary parts of each four-cell Bloch eigenvector are chosen as two independent real modes. Firstly, we match the eigenvalues from two calculations. Then, we compare isolated eigenvectors and nearly degenerate eigensubspaces using Eqs.~\ref{eq:single_mode_overlap} and~\ref{eq:mode_subspace_overlap}, respectively. For both carrier-off and carrier-on results, the reconstructed modes reproduce the same eigenvalues and eigenspaces as the original modes. Moreover, as expected, the reconstructed modes at $q_zc=0$ correspond to the unchanged original modes in Fig.~\figsubref{fig:three texture and corresponding}{d}, while those at $q_zc=\pm\pi/2$ correspond to the changed original modes. This agreement in both eigenvalues and eigenspaces confirms the reliability of our reconstruction, also suggesting the axial selectivity of the $g$-wave carrier contribution.

This resulting axial selectivity of the $g$-wave carrier contribution can be understood from Eqs.~\ref{eq:g-wave carrier} and~\ref{eq:reconstructed matrix}. 
Following the Hamiltonian separation in Eq.~\ref{eq:hamiltonian_carrier_split}, we decompose the single-cell Hessian into $H^q=H_a^q+\gamma_gH_g^q$, where the carrier contribution $H_g^q$ takes the form
\begin{equation}
    H^q_g=B_0^{(g)}+e^{iq_zc}B_{+1}^{(g)}+e^{-iq_zc}B_{-1}^{(g)},
    \label{eq:reconstructed matrix of carrier}
\end{equation}
where $B_0^{(g)}$ and $B_{\pm1}^{(g)}$ are the contributions of $\mathcal{H}_g$ to the corresponding matrices. As shown in Eq.~\ref{eq:g-wave carrier}, the carrier couples only neighboring cells, with opposite signs for the paired bonds at $+c$ and $-c$. This coupling pattern is also reflected in the carrier contribution to the Hessian. Accordingly, the extracted matrices satisfy $B_0^{(g)}=0$ and $B_{+1}^{(g)}=-B_{-1}^{(g)}$. As a result, Equation~\ref{eq:reconstructed matrix of carrier} reduces to
\begin{equation}
    H^q_g=(e^{iq_zc}-e^{-iq_zc})B_{+1}^{(g)}=2i\sin(q_zc)B_{+1}^{(g)},
    \label{eq:H_g with qzc}
\end{equation} 
which indicates that $H_g^q$ vanishes at $q_zc=0,\pi$, leaving the corresponding Hessian modes unchanged even when the carrier is switched on with finite $\gamma_g$. In contrast, $H_g^q$ is nonzero away from these nodes, allowing the carrier to modify the corresponding Hessian modes. Overall, these results indicate the axial selectivity of the $g$-wave carrier: it can modify Hessian modes only at nonnodal $q_z$ while having no effect on the nodal modes.

\subsection{Third-Harmonic Angular Selectivity of the Carrier}

\begin{figure*}[htbp]
    \centering
    \includegraphics[width=1\linewidth]{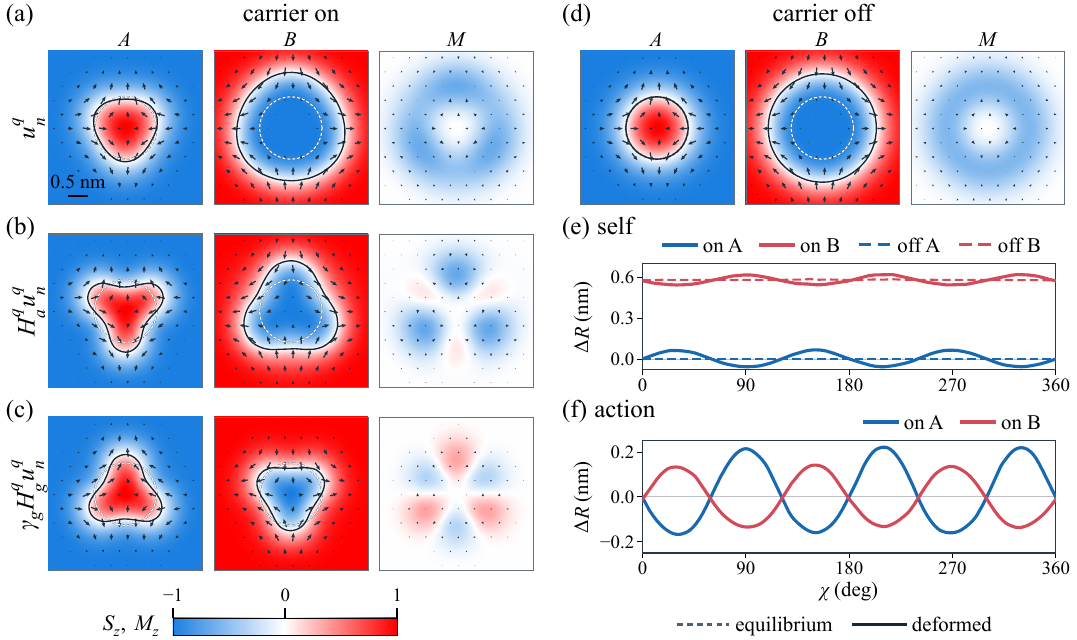}
    \caption{Reconstructed breathing-like mode and its Hessian actions at $q_zc=\pi/2$ in the second unit cell. (a)--(c) Spin deformations generated from the carrier-on eigenvector $u_n^q$, $H_a^q u_n^q$, and $\gamma_gH_g^q u_n^q$, respectively. (d) Corresponding carrier-off deformation. Columns show sublattices $A$, $B$ and the spin field $M$. Dotted and solid contours mark the equilibrium and deformed $S_z=0$ boundaries. (e), (f) Radial contour displacements $\Delta R(\chi)$ as functions of the azimuthal angle $\chi$ for the modes themselves in (a), (d) and the carrier action in (c), respectively.}
    \label{fig:self and action}
\end{figure*}

With the axial selectivity of the $g$-wave carrier established, we next investigate the modification of the carrier in the eigenmodes at nonnodal $q_z$ beyond the shift of eigenvalues. Indeed, its effect is already observable in the spin deformation of the original eigenmode 245. The deformations in Fig.~\figsubref{fig:two modes with on/off}{c} are nearly circular with the carrier off, while those in Fig.~\figsubref{fig:two modes with on/off}{d} exhibit a weak threefold modulation with the carrier on. 

To characterize this angular modulation, we investigate the reconstructed breathing-like mode at $q_zc=\pi/2$ corresponding to the original mode 245, and focus on the contribution of the carrier. We evaluate the contributions $H_a^qu_n^q$ and $\gamma_gH_g^qu_n^q$ separately, with the carrier-on eigenvector $u_n^q$. Their sum satisfies $H_a^qu_n^q+\gamma_gH_g^qu_n^q=\lambda_n^qu_n^q$, reproducing the mode itself.  Figures~\figsubref{fig:self and action}{a}--\figsubref{fig:self and action}{c} show the spin deformations generated from $u_n^q$, $H_a^qu_n^q$, and $\gamma_gH_g^qu_n^q$, respectively, in the second unit cell. The full four-cell deformations are shown in Fig.~S2 of the SM~\cite{SupplementaryMaterial}. The deformations in Figs.~\figsubref{fig:self and action}{a} and \figsubref{fig:self and action}{c} correspond to the mode itself and the carrier action, respectively. Meanwhile, Figure~\figsubref{fig:self and action}{d} shows the deformations associated with the carrier-off mode for comparison. The columns display the spin deformations on sublattices $A$ and $B$, together with the net spin field $M=(S_A+S_B)/2$. Considering that the equilibrium spin fields on the two sublattices cancel, $M$ directly displays the net spin change associated with the eigenmodes. Within each sublattice, the deformations generated from $H_a^qu_n^q$ and $\gamma_gH_g^qu_n^q$ exhibit threefold modulations, with opposite signs for these two contributions. Compared with the nearly circular carrier-off deformation in Fig.~\figsubref{fig:self and action}{d}, the carrier-on mode itself also exhibits a threefold angular modulation in Fig.~\figsubref{fig:self and action}{a}.

These angular modulations are further quantified by the radial displacement $\Delta R(\chi)$ of the $S_z=0$ contour relative to its equilibrium position. In Fig.~\figsubref{fig:self and action}{e}, the carrier-on profiles exhibit a threefold angular modulation, while the carrier-off ones remain nearly independent of $\chi$. The nearly constant carrier-off displacement corresponds to the approximately isotropic deformation in Fig.~\figsubref{fig:self and action}{d}, while the carrier introduces an additional threefold modulation. Moreover, the carrier-action profiles in Fig.~\figsubref{fig:self and action}{f} display the same threefold periodicity but with different amplitudes. The difference between mode-self and carrier-action profiles arises from the reduction of the opposite threefold contribution of $H_a^qu_n^q$ and $\gamma_gH_g^qu_n^q$. Besides, as the deformations are obtained from the real parts of the complex mode at a fixed phase, the relative axial phase between the sublattices $A$ and $B$ can produce different radial displacements on the two sublattices in Fig.~\ref{fig:self and action}. Overall, the results reveal a threefold in-plane angular selectivity in the breathing-like eigenmode.

The threefold modulation of the breathing-like mode raises the question of whether similar angular selectivity extends to other modes. However, identifying higher-order or mixed angular components directly from the deformation patterns and contour profiles as above may be difficult. Therefore, we introduce the azimuthal number $m$, which is widely used to characterize the in-plane spatial structure of skyrmion eigenmodes~\cite{2014_Internal_modes_of_a_skyrmion_for_mode_angular,2018_Localized_spin_waves,2018_Spin_eigenmodes_of_magnetic_skyrmions_for_mode_angular,2019_Kelvin_modes_of_a_skyrmion_line_for_qz,2019_Spin_eigenexcitations_of_an_antiferromagnetic_skyrmion_for_mode_angular}. This azimuthal number corresponds to the angular dependence $e^{im\chi}$. Using the net spin perturbation $\delta M=(\delta S_A+\delta S_B)/2$, we resolve its angular components and calculate their magnitudes as
\begin{equation}
    U_M(m)=\|\sum_i\mathbf \delta M_i^{\mathrm{tan}}e^{-im\chi_i}\|.
    \label{eq:angular spectrum}
\end{equation}
Here $\delta M_i^{\mathrm{tan}}$ contains the two local tangent components of the net spin perturbation represented by the $M$ channel in Fig.~\ref{fig:self and action}. $\|\cdot\|$ denotes their combined magnitude, and $\chi_i$ is the azimuthal angle at site $i$ relative to the skyrmion center. The sum runs over all the in-plane lattice sites. The same decomposition is applied to the eigenmode itself, $u_n^q$, and the carrier action, $\gamma_gH_g^qu_n^q$, providing an angular analysis of the $M$ channel in Figs.~\figsubref{fig:self and action}{a} and \figsubref{fig:self and action}{c}. The implementation details are provided in Sec.~S3 of the SM~\cite{SupplementaryMaterial}.

\begin{figure}[htbp]
    \centering
    \includegraphics[width=1.0\linewidth]{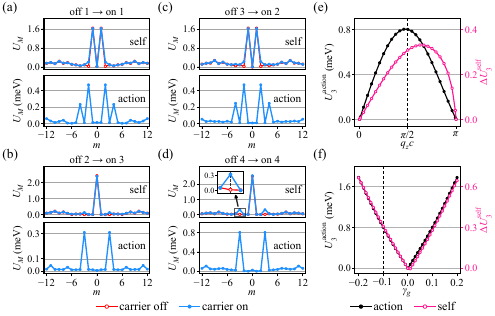}
    \caption{Angular spectra of Hessian modes themselves and their carrier actions, and third-harmonic amplitudes as functions of $q_z$ and $\gamma_g$. (a)--(d) Spectra for the first four carrier-off modes and their matched carrier-on modes at $q_zc=\pi/2$. The overlaps of carrier-off and carrier-on modes here all exceed $P=0.996$. The upper plots compare the carrier-off and carrier-on mode-self spectra, while the lower plots show the carrier action on the carrier-on modes. The inset in (d) shows an enlarged view around $m=-3$. (e) $U_3^{\mathrm{action}}$ and $\Delta U_3^{\mathrm{self}}$ as functions of $q_zc$ at $\gamma_g=-0.10$. (f) Corresponding dependence on $\gamma_g$ at $q_zc=\pi/2$.}
    \label{fig:angular and scanning}
\end{figure}

In Fig.~\figsubref{fig:angular and scanning}{d}, the angular spectra of the breathing-like mode are consistent with the spin deformations in Fig.~\ref{fig:self and action}. The carrier-off and carrier-on spectra of the mode itself largely overlap apart from $m=\pm3$, where $U_M(m=\pm3)$ increases markedly from $0.0180$ to $0.3266$ when the carrier is switched on. Meanwhile, the carrier-action spectrum from $\gamma_gH_g^qu_n^q$ is dominated by the $m=\pm3$ components, the same angular orders enhanced in the mode-self spectra. A similar selectivity appears in Fig.~\figsubref{fig:angular and scanning}{b}, where the mode itself is predominantly $m=0$ and the carrier action is dominated by $m=\pm3$. A further correspondence emerges in Figs.~\figsubref{fig:angular and scanning}{a} and \figsubref{fig:angular and scanning}{c}, where modes dominated by $m=\pm1$ have carrier-action spectra with main peaks at $m=\pm2$ and $\pm4$. Weaker changes at these angular orders are also visible in the self spectra, while the dominant $m=\pm1$ components remain nearly unchanged.

Figures~\figsubref{fig:angular and scanning}{a}--\figsubref{fig:angular and scanning}{d} together reveal a common selective shift of three in angular order between the modes themselves and their carrier actions. Given that the spectra are symmetric under $m\rightarrow-m$, this selectivity can be expressed in terms of nonnegative angular orders as 
\begin{equation}
    m_{\mathrm{action}}=|m_{\mathrm{self}}\pm3|.
    \label{eq:three-order shift}    
\end{equation}
Therefore, the angular structure of the carrier action is directly related to the components already present in the mode itself. Taking Fig.~\figsubref{fig:angular and scanning}{a} as an example, a component in the mode itself with angular order $m=1$ gives $m-3=-2$ and $m+3=4$. This result corresponds to the nonnegative orders 2 and 4 observed in the carrier action. For modes with mixed angular components, the relation in Eq.~\ref{eq:three-order shift} applies to all main components and produces several corresponding peaks in the action spectrum. The prominent action peaks, in turn, reflect the selective modification of the angular structure of the modes themselves. Thus, the action spectra link the angular content already present in the modes themselves to the angular components modified by the carrier. Remarkably, even in the presence of mixed angular components, the same relation extends to all the lowest 12 modes, as shown in Fig.~S3 of the SM~\cite{SupplementaryMaterial}. Taken together, these results establish a common $\Delta m=\pm3$ angular selectivity across the examined modes.

The axial dependence and parameter robustness of this in-plane angular modification are further explored. We take the third-harmonic modification of the breathing-like mode in Fig.~\figsubref{fig:angular and scanning}{d} as a representative example. This modification is tracked both through the third-harmonic amplitude in the carrier-action spectrum, $U_3^{\mathrm{action}}$, and the corresponding change in the mode-self spectrum, $\Delta U_3^{\mathrm{self}}$. In Fig.~\figsubref{fig:angular and scanning}{e}, both quantities rise to a maximum and then decrease as $q_zc$ increases, vanishing at the axial nodes $q_zc=0,\pi$. Meanwhile, $U_3^{\mathrm{action}}$ exhibits an approximately sinusoidal dependence on $q_zc$, consistent with the $\sin(q_zc)$ factor in Eq.~\ref{eq:H_g with qzc}. The different peak positions suggest that the third-harmonic change in the mode does not simply follow the magnitude of the carrier action, as the mode is determined by both $H_a^q$ and $\gamma_gH_g^q$. As shown in Fig.~\figsubref{fig:angular and scanning}{f}, the third-harmonic modification is observed for both signs of $\gamma_g$, with the magnitudes of both quantities generally increasing with $|\gamma_g|$. These results show the dependence of the axial wave vector and carrier strength on the angular modification, and demonstrate its robustness over a wide range of $\gamma_g$.

\section{Discussion}

The axial and in-plane angular selectivity discussed above can be related to the bulk $g$-wave symmetry of \ch{CrSb}. In this $g$-wave altermagnet, characteristic $g$-wave patterns have been demonstrated in electronic spin splitting~\cite{2026_3D_bulk-resolved_g-wave_altermagnetic_order_parameter_in_CrSb_for_g_spin_splitting,2024_Large_Band_Splitting_in_g-Wave_Altermagnet_CrSb_for_g_spin_splitting} and magnon splitting~\cite{g-wave_model1}. Similar patterns also appear in electronic spin splitting of other $g$-wave altermagnets, including $\mathrm{Fe}_{1/4}\mathrm{NbS}_2$ and $\mathrm{V}_{1/3}\mathrm{NbS}_2$~\cite{g-wave_model2}. For this $g$-wave symmetry, the leading-order form factor of $g$-wave splitting can be represented as~\cite{2024_Minimal_models_for_altermagnetism}
\begin{equation}
    f_g(k)=k_zk_y(3k_x^2-k_y^2)=k_zk_\perp^3\sin(3\chi_{k}),
    \label{eq:g-wave form factor}
\end{equation}
where $k_\perp=\sqrt{k_x^2+k_y^2}$ and $\chi_{k}$ is the in-plane azimuthal angle of $k$. This form combines the axial factor $k_z$ with the in-plane angular third-harmonic factor $\sin(3\chi_{k})$. 

In our model, the $g$-wave carrier produces both the axial dependence described by $\sin(q_zc)$ and the angular order shift $\Delta m=\pm3$. In the long-wavelength limit, the axial factor $\sin(q_zc)$ reduces to a term linear in $q_z$, consistent with the factor $k_z$ in Eq.~\ref{eq:g-wave form factor}. Besides, the in-plane factor $\sin(3\chi_k)$ in the splitting reflects the $g$-wave AM symmetry encoded in the carrier. This suggests that the carrier acts on momentum-space components through a similar third-harmonic angular factor. Therefore, when acting on a component with angular dependence $e^{im\chi_k}$, the carrier generates components of orders $m+3$ and $m-3$. Given that the Fourier transform preserves the angular index of a harmonic, the resulting $m\pm3$ components retain their angular orders in real space. This provides a symmetry-based interpretation of the $\Delta m=\pm3$ shift observed in our real-space angular spectra in Fig.~\ref{fig:angular and scanning}. 

The breathing-like mode in Fig.~\ref{fig:self and action} provides a direct illustration of the in-plane angular selectivity. Its mode-self spectrum has a single dominant angular order $m=0$, providing an approximately angle-independent reference to identify the modulation induced by the carrier. Hence, the third-harmonic modulation induced by the carrier is particularly clear in the $M_z$ channel of its carrier action in Fig.~\figsubref{fig:self and action}{c}, where three positive regions alternate with three negative ones. This in-plane structure closely resembles the angular pattern in both electronic splitting and magnon splitting~\cite{g-wave_model2,g-wave_model1}. Overall, these results connect the bulk $g$-wave symmetry to the angular modification of fluctuations in the skyrmion tubes, offering another real-space manifestation of this symmetry.

Beyond the Hessian eigenmodes, the axial and in-plane angular selectivity also extends to the dynamic eigenmodes. At wave vector $q_z$, the dynamic eigenmodes can be obtained from the linearized Landau-Lifshitz-Gilbert equation~\cite{2018_Localized_spin_waves,2023_Eigenmodes_of_magnetic_skyrmion_lattices_for_dynamics_modes,2020_Spin_waves_in_skyrmionic}
\begin{equation}
    \partial_t \delta S^q=-i\sigma^yH^q\delta S^q=D^q\delta S^q,
    \label{eq:dynamics}
\end{equation}
where $\sigma^y=\begin{pmatrix}
    0 & -iI_s \\
    iI_s & 0
\end{pmatrix}$ is the Pauli matrix with identity matrix $I_s$. $\delta S^q$ denotes the single-cell counterpart of $\delta S$ at a given $q_z$. In this equation, we omit the overall prefactor $\gamma/\mu_s$ and set the damping to zero for convenience. Following the Hessian separation, we also decompose the dynamic matrix $D^q$ into
\begin{equation}
    D^q=D^q_a+D^q_g=-i\sigma^y(H^q_a+\gamma_gH^q_g),
    \label{eq:dynamics_separation}
\end{equation}
where $D^q_a$ and $D^q_g$ denote the conventional and $g$-wave carrier contributions, respectively. As $H^q_g$ in Eq.~\ref{eq:H_g with qzc} vanishes at $q_zc=0,\pi$, the carrier contribution $D^q_g$ also vanishes at these nodes. Thus, the corresponding dynamic modes are unaffected by the carrier. Moreover, we find that the breathing-like dynamic mode also exhibits threefold modulation at $q_zc=\pi/2$, as shown in Sec.~S4 of the SM~\cite{SupplementaryMaterial}. Therefore, the axial and in-plane angular selectivity is inherited by the dynamic modes.

Recent experiments have demonstrated spectroscopic access to both the $q_z$-dependent dynamics and modes with distinct in-plane angular character. $q_z$-dependent dynamics have been investigated using propagating spin-wave spectroscopy, polarized inelastic neutron scattering, and neutron spin-echo spectroscopy~\cite{2020_Propagation_dynamics_of_spin_excitations,2022_Topological_magnon_band_structure,2023_Asymmetric_slow_dynamics}. More recently, Brillouin light scattering (BLS) microscopy has identified a quadrupole mode ($m=2$) and possibly a sextupole mode ($m=3$)~\cite{2025_Short-wave_magnons_with_multipole}, suggesting that BLS could probe the third azimuthal harmonic of the breathing-like mode predicted in our work. Taken together, these experimental advances open up possible avenues for probing the $q_z$-dependent in-plane angular selectivity, which may serve as an experimental signature of bulk $g$-wave AM symmetry in skyrmion tubes.

\section{Conclusion}

In summary, we have investigated the influence of bulk $g$-wave altermagnetic symmetry on localized skyrmion tubes within a frustrated atomistic model. We show that the axial cancellation of the $g$-wave carrier contribution leaves the static textures and energy unchanged, while the energy curvature remains sensitive to the carrier. Indeed, the carrier contribution to the Hessian is proportional to $\sin(q_zc)$, which vanishes at $q_zc=0,\pi$ and accounts for the selective modification along the tube axis. A characteristic third-harmonic angular shift, $\Delta m=\pm3$, is revealed in the angular spectra, corresponding to the observed in-plane angular selectivity. This angular selectivity is particularly clear in the breathing-like mode, providing a new real-space manifestation of altermagnetic symmetry. Furthermore, the $q_z$-dependent in-plane angular selectivity extends to the dynamic eigenmodes and may serve as an experimental signature of bulk $g$-wave AM in skyrmion tubes. Our results provide new insight into bulk $g$-wave altermagnetism and its influence on the skyrmion tubes.

\ack{This work was supported by the National Key R\&D Program of China (Grant No.~2024YFA1408303, 2022YFA1403301), the National Natural Sciences Foundation of China (Grants No.~12474247 and 92165204). Yusheng Hou acknowledges the support from Guangdong Provincial Key Laboratory of Magnetoelectric Physics and Devices (Grant No.~2026B1212170049) and Research Center for Magnetoelectric Physics of Guangdong Province (Grant No.~2024B0303390001). Numerical calculations are performed on Tianhe-Xingyi.}

\data{The data that support the findings of this study are available from the authors upon reasonable request.}

\FloatBarrier
\bibliographystyle{ropp_compact}
\bibliography{reference}

\end{document}